\documentclass[reprint,aps,prb,groupedaddress,floatfix]{revtex4-2}

\usepackage{physics}
\usepackage{amsmath}
\usepackage{xcolor}
\usepackage{graphicx}
\usepackage[hidelinks]{hyperref}
\usepackage{cleveref}
\usepackage{float}
\usepackage[normalem]{ulem} % Required for \sout
\crefname{equation}{Eq.}{} % For multiple \eqrefs needed.

\newcommand{\pati}[2]{}

\begin{document}

\title{Scattering at Space-Time Interfaces \\ between Dispersive Media}

\author{Klaas De Kinder}
\author{Christophe Caloz}
\email[]{christophe.caloz@kuleuven.be}
\affiliation{Department of Electrical Engineering, KU Leuven, Leuven, 3000, Belgium}

\date{\today}

\begin{abstract}
    Dynamic modulation of material properties in space and time enables powerful control over wave propagation, yet existing theories largely rely on idealized, nondispersive models. In realistic media, frequency dispersion can strongly reshape wave dynamics, especially near resonances in highly dispersive platforms such as epsilon-near-zero materials. Here, we develop a general frequency transition theory for electromagnetic scattering at moving interfaces between dispersive media. From phase continuity, we derive nonlinear frequency transition relations and show that dispersion fundamentally reshapes the space-time scattering landscape, enabling additional propagating solutions with no counterpart in nondispersive systems. Applied to Drude, Lorentz and double-Drude media, the theory reveals how resonant dispersion, material loss and negative-index branches reorganize the scattering channels. For the two-wave scattering class, we further introduce a mixed-domain formulation that combines time-domain interface kinematics with frequency-domain constitutive relations, yielding closed-form scattering coefficients. These results establish a unified framework for dispersive space-time scattering and open opportunities for dispersion-based transition engineering in realistic materials.
\end{abstract}
    
\maketitle

\section{Introduction}
    \pati{Space-Time Varying Media}{
    }
    
    Dynamic modulation of material parameters, such as the refractive index, in both space and time has emerged as a powerful paradigm for controlling wave phenomena~\cite{Cassedy1963_PUB,Cassedy1967_PUB,Caloz2019a_ST_Metamaterials_USTEM_PUB,Caloz2019b_ST_Metamaterials_USTEM_PUB}. These modulations, typically induced by external electronic, optical, acoustic or mechanical drives~\cite{Saleh2019_BOOK,Rhodes1981_Signal_Proc_PUB,Shalaev2019_Active_Meta_PUB}, enable simultaneous frequency and wave-vector conversion, thereby extending wave manipulation beyond the limits of static media~\cite{Caloz2022_GSTEMs_ASTEM_PUB}. This capability has led to a wide range of novel effects and applications, including temporal analogs of Faraday rotation~\cite{Alu2022_Nonreci_Faraday_TEM_PUB,Huanan2023_Faraday_TEM_PUB}, arbitrary pulse shaping~\cite{Bahrami2025_Pulse_Shap_ASTEM_PUB}, interface-based light amplification~\cite{Pendry2021_Gain_TEM_PUB,Luo2023_Param_Amplf_USTEM_PUB}, generalized frequency chirping~\cite{DeKinder2026_Scat_Chirp_ASTEM_PUB}, temporal aiming~\cite{Engheta2020_Aiming_TEM_PUB}, space-time reversal~\cite{Fink2016_TR_TEM_PUB}, temporal beam splitting~\cite{Guerreiro2003_Split_Interf_TEM_PUB}, motion-induced photon cooling~\cite{Pendry2024_Air_Cond_Phot_USTEM_PUB} and Doppler pulse amplification~\cite{DeKinder2025_DoPA_USTEM_PUB}.
    
    \pati{Importance of Dispersion}{
    }

    Despite these advances, most existing studies rely on idealized, nondispersive media. In practice, however, all physical materials are inherently dispersive due to causality, leading to frequency-dependent responses that may significantly affect wave propagation~\cite{Saleh2019_BOOK}. The impact of dispersion is particularly pronounced in highly dispersive platforms, such as epsilon-near-zero materials, where rapid variations of the refractive index occur near resonances and underpin recent experiments~\cite{Shalaev2023_PTC_TEM_PUB,Boyd2020_Time_Refr_ENZ_TEM_PUB,Segev2023_Single_Cycle_TEM_PUB,Kinsey2025_ST_Knife_TEM_PUB}. Beyond being a constraint, dispersion also creates new opportunities. In particular, it enables dispersion-based transition engineering, where the accessible scattering states are controlled by the interplay between material dispersion and interface velocity.

    \pati{Literature Dispersive TEM}{
    }
    
    Recent work has begun to address dispersive \emph{time}-varying media~\cite{Fante1971_Transmission_TEM_PUB,Pendry2022_Review_TEM_PUB,Mirmoosa2022_Dipole_TEM_PUB,Joannopoulos2024_Opt_Prop_TEM_PUB,Maslov2021_Const_Rel_PUB,Maslov2024_Dispersion_PUB,Monticone2022_Dispersion_TEM_PUB}, establishing fundamental constraints such as Kramers-Kronig relations for temporal media and exploring their implications for wave dynamics~\cite{Monticone2021_Spectral_Caus_Dispersion_TEM_PUB,Engheta2021_Kramers_Kronig_TEM_PUB}. Wave scattering has been investigated for several canonical temporal dispersive models, including Drude~\cite{Horsley2023_Eigenpulses_TEM_PUB}, Lorentz~\cite{Engheta2021_Dispersion_TEM_PUB,Maslov2021_Light_Scat_Dispersion_TEM_PUB,Horsley2025_Scat_TEM_Drude_Lorentz_TEM}, double-Drude~\cite{Sirota2023_Neg_Ref_TEM_PUB} and hyperbolic media~\cite{Engheta2025_Disp_Hyper_Media_TEM_PUB}. In parallel, modal descriptions such as temporal quasi-normal modes have been introduced to characterize these systems~\cite{Horsley2025_Quasi_Normal_Mode_TEM}.
    
    \pati{Gap Dispersive USTEM}{
    }

    In contrast, dispersive \emph{space-time} varying systems remain virtually unexplored~\cite{Felsen1972_Pulse_Prop_Dispersion_USTEM_PUB}. Existing studies~\cite{Gaafar2019_PUB,Agrawal2015_Temporal_Analog_USTEM_PUB,Agrawal2016_Spectral_Splitting_USTEM_PUB,Agrawal2021_Refl_and_Trans_USTEM_PUB,Wang2024_Soliton_Disp_USTEM_PUB} are primarily restricted to pump-probe configurations in dispersive nonlinear optical fibers and rely on several limiting assumptions. First, dispersion is typically approximated through a Taylor expansion, restricting the analysis to frequencies far from material resonances. Second, the slowly varying envelope approximation is employed, limiting the treatment to narrow-band signals. Third, frame transformations to a purely temporal reference frame are used, which implicitly assume a single constant modulation velocity and preclude accelerated or multi-velocity interfaces. Finally, the class of material modulations considered is often restricted, limiting applicability across general electromagnetic responses. Taken together, these constraints prevent the development of a general scattering framework. A complete theory for wave scattering at interfaces in dispersive space-time varying media is therefore still lacking.

    \pati{Contribution}{
    }

    In this work, we develop a generalized scattering theory for wave interaction with moving interfaces separating two dispersive media. We show that dispersion fundamentally reshapes the space-time scattering landscape by enabling additional propagating solutions with no counterpart in nondispersive systems, which we term ``dispersion-mediated space-time modes''. The proposed framework accommodates arbitrary dispersion profiles, including resonant regimes where Taylor expansions break down. The analysis is carried out entirely in the laboratory frame, avoiding frame transformations and therefore remaining compatible with multiple modulation velocities and accelerated interfaces. In addition to the general frequency transition theory, we derive closed-form scattering coefficients for the two-wave scattering class by introducing a mixed-domain formulation. By explicitly incorporating material dispersion into the scattering process, this work establishes a rigorous foundation for wave interactions in realistic space-time varying media.

    \pati{Organization}{
    }

    The paper is structured as follows. Section~\ref{sec:Frequency_Transitions} develops the theory of frequency transitions for scattering at space-time interfaces between dispersive media and presents applications to Drude, Lorentz and double-Drude models. Section~\ref{sec:Scattering_Coefficients} derives the corresponding scattering coefficients for the two-wave class and applies the formulation to the Drude model. Section~\ref{sec:Comparison_with_Nondispersive_Media} compares the dispersive results with their nondispersive counterparts and validates the theory with a full-wave simulation. Section~\ref{sec:Experimental_Outlook} outlines potential experimental realizations of the dispersion-mediated space-time modes. Finally, Sec.~\ref{sec:Conclusions} concludes the paper and outlines future research directions.

\section{Frequency Transitions}\label{sec:Frequency_Transitions}
    \subsection{Theory}\label{subsec:Theory_Freq_Trans}
        \pati{Assumptions}{
        }

        We consider one-dimensional electromagnetic scattering at a space-time interface moving with constant velocity $v_{\text{m}}$, which may be positive (codirectional) or negative (contradirectional). The formulation includes the limiting cases $v_{\text{m}} = 0$ (stationary interface) and $v_{\text{m}} \rightarrow \infty$ (pure-time interface). The interface separates two dispersive media, each characterized by a frequency-dependent refractive index $n_{i}{\left[\omega\right]}$~\footnote{Spatial dispersion typically arises from the motion of matter constituents, which induce effective bianisotropy in uniformly moving media~\cite{Rontgen1888_PUB,Kong2008_Wave_Theory_BOOK}. Here, however, the modulation is a prescribed spatiotemporal variation---or perturbation---of the electromagnetic material parameters, without physical transport of matter. The media therefore remain at rest, and no nonlocal constitutive effects are introduced beyond the assumed temporal frequency dispersion.}, where $i=1,2$ denotes the medium, satisfying the dispersion relation
        \begin{equation}\label{eq:Dispersion_Relation}
            k_{i}^{2}{\left[\omega\right]} = n_{i}^{2}{\left[\omega\right]}\frac{\omega^{2}}{c^{2}}\,.
        \end{equation}
        An incident wave with frequency $\omega_{\text{i}}$ in the first medium impinges on the moving interface and generates scattered waves with shifted frequencies in both media.        

\begin{figure*}
    \centering
    \includegraphics[width=\linewidth]{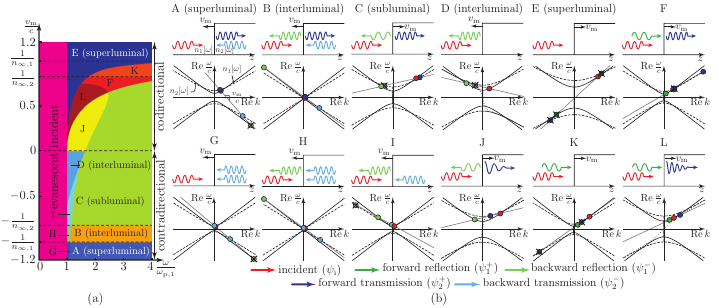}
    \caption{Dispersion-mediated space-time modes in a lossless Drude medium [Eq.~\eqref{eq:Refractive_Index_Drude_Media}], with $n_{\infty,1} = 1$, $\omega_{\text{p},1} = 5$, $n_{\infty,2} = 1.22$ and $\omega_{\text{p},2} = 10$. (a)~Scattering regimes as a function of the modulation velocity, $v_{\text{m}}$, and the incident frequency, $\omega_{\text{i}}$. The dashed lines indicate the corresponding nondispersive velocity limits. (b)~Space-index representation of the different space-time states with associated spectral transition diagrams, where some of the solutions are crossed out because they do not satisfy the selection conditions in Eqs.~\eqref{eq:Selection_Conditions}. The dotted curves represent dispersion relations evaluated at fixed values of the imaginary part of the frequency solutions.}
    \label{fig:Number_of_Waves_Drude}
\end{figure*}

        \pati{Phase Matching Method}{
        }

        The scattered frequencies are obtained by enforcing phase continuity between the incident and scattered fields at the interaction points along the interface trajectory $z=v_{\text{m}}t$. Using the time-harmonic convention $\exp\left(i\left(k{\left[\omega\right]}z-\omega t\right)\right)$, phase continuity yields
        \begin{subequations}\label{eq:Phase_Matching_Conditions}
            \begin{align}
                k_{1}{\left[\omega_{1}\right]}z - \omega_{1}t &= \left.k_{1}{\left[\omega_{\text{i}}\right]}z - \omega_{\text{i}}t\right|_{z=v_{\text{m}}t}\,, \\
                k_{2}{\left[\omega_{2}\right]}z - \omega_{2}t &= \left.k_{1}{\left[\omega_{\text{i}}\right]}z - \omega_{\text{i}}t\right|_{z=v_{\text{m}}t}\,,
            \end{align}
        \end{subequations}
        where $\omega_{1}$ and $\omega_{2}$ denote the reflected and transmitted frequencies, respectively. Inserting Eq.~\eqref{eq:Dispersion_Relation}, the trajectory parametrization and eliminating the explicit time dependence in Eqs.~\eqref{eq:Phase_Matching_Conditions}, we obtain the characteristic equations governing the frequency transitions:
        \begin{subequations}\label{eq:General_Characteristic_Equations}
            \begin{align}
                \left(1+n_{1}{\left[\omega_{1}\right]}\frac{v_{\text{m}}}{c}\right)\omega_{1} &= \left(1-n_{1}{\left[\omega_{\text{i}}\right]}\frac{v_{\text{m}}}{c}\right)\omega_{\text{i}}\,, \\
                \left(1-n_{2}{\left[\omega_{2}\right]}\frac{v_{\text{m}}}{c}\right)\omega_{2} &= \left(1-n_{1}{\left[\omega_{\text{i}}\right]}\frac{v_{\text{m}}}{c}\right)\omega_{\text{i}}\,.
            \end{align}
        \end{subequations}
        In a dispersion diagram, the solutions to Eqs.~\eqref{eq:General_Characteristic_Equations} correspond to the intersections between the dispersion curves, defined by Eq.~\eqref{eq:Dispersion_Relation}, and a transition line of slope~$v_{\text{m}}$ passing through the incident point~$\left(k_{1}{\left[\omega_{\text{i}}\right]},\omega_{\text{i}}/c,\right)$. In the nondispersive limit, Eqs.~\eqref{eq:General_Characteristic_Equations} reduce to the standard Doppler relations~\cite{Caloz2019b_ST_Metamaterials_USTEM_PUB}. In dispersive media, however, Eqs.~\eqref{eq:General_Characteristic_Equations} are nonlinear, so that multiple solutions may arise, each corresponding to a distinct scattered mode. Therefore, we denote the solutions to Eqs.~\eqref{eq:General_Characteristic_Equations} in medium~$i$ by~$\omega_{i}^{\pm}$, where the superscripts indicate forward~($+$) and backward~($-$) propagation.
        
        \pati{Physical Wave Conditions}{
        }

        Not all mathematical solutions of Eqs.~\eqref{eq:General_Characteristic_Equations} correspond to physically admissible waves. Additional selection criteria are therefore required. We introduce two conditions: a kinematic constraint based on group velocity and an energetic constraint based on media passivity. The kinematic condition requires that a scattered wave must not intersect the interface trajectory after the interaction. This requirement is enforced by comparing the interface trajectory $z=v_{\text{m}}t$ with the wave-packet trajectory $z=v_{\text{g},i}{\left[\omega_{i}^{\pm}\right]}t$, where $v_{\text{g},i}{\left[\omega_{i}^{\pm}\right]} = \partial \omega_{i}^{\pm}/\partial k_{i}{\left[\omega_{i}^{\pm}\right]}$ is the group velocity of wave $\omega_{i}^{\pm}$. For transmitted waves in the second medium, the wave packet must move ahead of the interface, i.e., $v_{\text{g},2}{\left[\omega_{2}^{\pm}\right]} > v_{\text{m}}$, whereas for reflected waves in the first medium, the wave packet must remain behind the interface, i.e., $v_{\text{g},1}{\left[\omega_{1}^{\pm}\right]} < v_{\text{m}}$. When the scattered wave and the interface propagate in opposite directions, these conditions are automatically satisfied since they separate immediately after interaction, for instance for a contradirectional moving interface and a forward transmitted wave. In contrast, when they propagate in the same direction, these conditions become nontrivial and must be explicitly enforced. This occurs, for instance, for a forward transmission with a codirectional interface. The second condition follows from media passivity. When the solutions $\omega_{i}^{\pm}$ and $k_{i}{\left[\omega_{i}^{\pm}\right]}$ are complex, the wave amplitude must not grow in a passive medium. Evaluating the field along its wave-packet trajectory $z=v_{\text{g},i}{\left[\omega_{i}^{\pm}\right]}t$, the amplitude evolves as $\exp\left(-\left(\Im k_{i}{\left[\omega_{i}^{\pm}\right]}\cdot v_{\text{g},i}{\left[\omega_{i}^{\pm}\right]} - \Im \omega_{i}^{\pm}\right)t\right)$. Passivity therefore requires that $\Im k_{i}{\left[\omega_{i}^{\pm}\right]}\cdot v_{\text{g},i}{\left[\omega_{i}^{\pm}\right]} - \Im \omega_{i}^{\pm} > 0$, which guarantees decay or, at most, bounded propagation. The two conditions, group-velocity restriction and media passivity, can be summarized as
        \begin{subequations}\label{eq:Selection_Conditions}
            \begin{align}
                v_{\text{g},1}{\left[\omega_{1}^{\pm}\right]} &< v_{\text{m}}\,,  \\
                v_{\text{g},2}{\left[\omega_{2}^{\pm}\right]} &> v_{\text{m}}\,, \label{eq:Group_Velocity_Restriction_Second_Medium} \\
                \Im k_{i}{\left[\omega_{i}^{\pm}\right]}\cdot v_{\text{g},i}{\left[\omega_{i}^{\pm}\right]} &> \Im\omega_{i}^{\pm}\,. \label{eq:Media_Passivity_Condition}
            \end{align}
        \end{subequations}

    \subsection{Application}\label{subsec:Application_Freq_Trans}
        \subsubsection{Drude Media}\label{subsubsec:Drude_Media}
            \pati{Specific Drude Solutions}{
            }

            We apply the general theory developed in the previous section to the lossless Drude model, in which the refractive index is given by~\cite{Saleh2019_BOOK,Fox2001_Opt_Prop_Solids_BOOK}
            \begin{equation}\label{eq:Refractive_Index_Drude_Media}
                n_{i}{\left[\omega\right]} = \sqrt{n_{\infty,i}^{2} - \frac{\omega_{\text{p},i}^{2}}{\omega^{2}}} \,
            \end{equation}
            where $n_{\infty,i}$ and $\omega_{\text{p},i}$ are the high-frequency refractive index and plasma frequency of medium $i=1,2$, respectively. Substituting Eq.~\eqref{eq:Refractive_Index_Drude_Media} into Eqs.~\eqref{eq:General_Characteristic_Equations} yields two quadratic equations for the reflected and transmitted frequencies, which can be solved analytically (see Sec.~\ref{subsec:appendix:Drude_Media}). This results in four candidate solutions, namely two reflected and two transmitted, which must be filtered using the physical admissibility conditions [Eqs.~\eqref{eq:Selection_Conditions}]. Depending on the incident frequency and the modulation velocity, different subsets of these solutions remain, leading to qualitatively distinct scattering regions. This rich solution structure gives rise to what we term ``dispersion-mediated space-time modes'', namely scattering configurations that have no counterpart in nondispersive problems. Figure~\ref{fig:Number_of_Waves_Drude} illustrates this landscape for the Drude model. Figure~\ref{fig:Number_of_Waves_Drude}a plots the scattering regions as a function of incident frequency and modulation velocity, while Fig.~\ref{fig:Number_of_Waves_Drude}b provides the corresponding space-index perspectives and spectral transition diagrams. In total, twelve distinct regimes, labeled~A through~L, are identified.

            \pati{Conventional Modes}{
            }

            We first identify the regimes that correspond to known nondispersive scattering behaviors. These are recovered in the high-frequency limit, where dispersion becomes negligible. Region~C corresponds to the subluminal regime, where an incident wave generates one backward reflected wave and one forward transmitted wave. A second candidate transmitted wave is excluded by the group-velocity condition, as its trajectory would cross the interface after scattering. Region~B corresponds to the contradirectional interluminal regime. In this case, multiple scattered waves are generated, including reflected and transmitted components, all of which satisfy the admissibility conditions. Region~D represents the codirectional interluminal regime, where only a reflected wave exists. Notably, in contrast to the nondispersive case, this regime can occur for contradirectional interfaces, illustrating how dispersion modifies the regime structure. Regions~A and~E correspond to superluminal regimes. In region~E (codirectional), no scattering occurs because the interface moves faster than the incident wave. In region~A (contradirectional), the incident wave generates two transmitted waves, one later-forward and one later-backward. A backward-propagating mode in the first medium is excluded by the group-velocity condition. More generally, dispersion fundamentally alters the regime boundaries. In nondispersive systems, these boundaries depend only on the modulation velocity. In contrast, here they become frequency dependent due to dispersion, leading to shifts and distortions of the regime map and, in some cases, extending regimes beyond their nondispersive limits.

            \pati{Regions~G,~H and~I}{
            }

            Beyond the conventional regimes, the Drude model supports purely dispersive space-time states with no nondispersive analog, corresponding to regions~F through~L. Regions~G,~H and~I occur for contradirectional interfaces and incident frequencies near the plasma frequency of the first medium. In region~I, the scattering consists of a backward reflected wave and a backward transmitted wave. No forward transmitted wave exists because the transition line does not intersect the positive-slope branch of the second medium. A second backward transmitted candidate is excluded by the group-velocity condition. As the magnitude of the modulation velocity decreases, i.e., as the transition line becomes more vertical, the system transitions into region~H. In this regime, the transition line intersects an additional branch of the dispersion relation, giving rise to a second backward transmitted wave that satisfies the admissibility conditions. Upon further reduction of the velocity (region~G), the reflected wave is eliminated by the group-velocity condition, as the interface overtakes its trajectory. A common feature of regions~G--I is that one of the transmitted waves exhibits a near-zero group velocity, corresponding to a nearly horizontal dispersion branch at the intersection point. As a result, this mode remains localized near its scattered position and behaves as a quasi-stationary state.

            \pati{Regions~F,~J,~K and~L}{
            }

            Regions~F,~J,~K and~L occur for codirectional interfaces and, unlike regions~G--I, are not restricted to frequencies near the plasma frequency. Region~J resembles an interluminal regime, but with a key difference: the forward transmitted wave is evanescent in the second medium and therefore propagates only over a finite distance. Regions~F,~K and~L exhibit forward-propagating modes in the first medium, arising from intersections of the transition line with positive-slope branches of the dispersion relation. The existence of such modes implies that their group velocity is smaller than the modulation velocity; otherwise, they would be excluded by the group-velocity condition. Depending on the specific parameters, the transmitted wave in these regimes may be propagating (region~F), evanescent (region~L), or entirely suppressed by the admissibility conditions (region~K).

\begin{figure}
    \centering
    \includegraphics[width=0.8\linewidth]{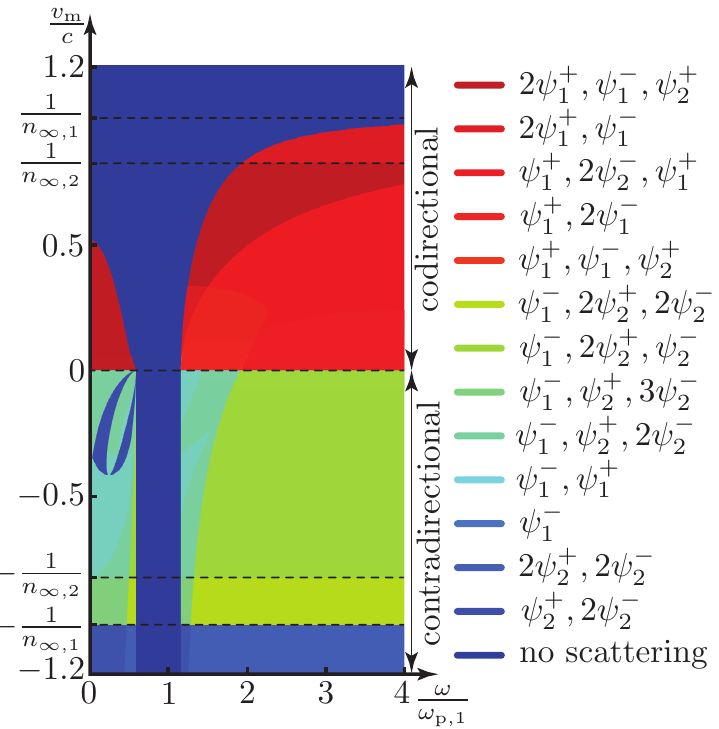}
    \caption{Dispersion-mediated space-time modes for the Lorentz dispersion model [Eq.~\eqref{eq:Refractive_Index_Lorentz_Media}] with parameters $n_{\infty,1} = 1$, $\omega_{\text{p},1} = 5$, $\omega_{0,1} = 3$, $\gamma_{1} = 0.001$, $n_{\infty,2} = 1.22$, $\omega_{\text{p},2} = 10$, $\omega_{0,2} = 5$ and $\gamma_{2} = 0.001$. The dashed lines indicate the nondispersive velocity limits.}
    \label{fig:Number_of_Waves_Lorentz}
\end{figure}

        \subsubsection{Lorentz Media}\label{subsubsec:Lorentz_Media}
            \pati{Lorentz Media}{
            }

            As a second example, we consider the Lorentz dispersion model, which provides a more realistic description of material response by incorporating both resonance and loss mechanisms. The refractive index is given by~\cite{Saleh2019_BOOK,Fox2001_Opt_Prop_Solids_BOOK}
            \begin{equation}\label{eq:Refractive_Index_Lorentz_Media}
                n_{i}{\left[\omega\right]} = \sqrt{n_{\infty,i}^{2} + \frac{\omega_{\text{p},i}^{2}}{\omega_{0,i}^{2}-i\gamma_{i}\omega -\omega^{2}}}\,,
            \end{equation}
            where $\omega_{0,i}$ and $\gamma_{i}$ denote the resonance frequency and material loss term of medium $i=1,2$, respectively. In contrast to the Drude model [Eq.~\eqref{eq:Refractive_Index_Drude_Media}], the Lorentz model introduces both a resonant pole and a finite linewidth, leading to a richer and more intricate dispersion relation. Substituting Eq.~\eqref{eq:Refractive_Index_Lorentz_Media} into Eqs.~\eqref{eq:General_Characteristic_Equations} yields quartic polynomial equations (see Sec.~\ref{subsec:appendix:Lorentz_Media}), allowing for up to four distinct solutions per medium. After application of the admissibility conditions in Eqs.~\eqref{eq:Selection_Conditions}, the corresponding space-time regime map shown in Fig.~\ref{fig:Number_of_Waves_Lorentz} contains up to fourteen distinct regimes, reflecting the larger number of resonant branches and their interaction with the transition line.

            Beyond this quantitative increase, the Lorentz model introduces qualitative differences. Near resonance, the dispersion curve exhibits strong curvature and may support both normal and anomalous dispersion. As a consequence, small changes in modulation velocity can produce large shifts in the allowed frequency transitions, and distinct roots may cluster close to one another. In addition, the damping parameter $\gamma_i$ makes both $\omega_i^{\pm}$ and $k_i{\left[\omega_i^{\pm}\right]}$ complex in general. The passivity condition in Eqs.~\eqref{eq:Media_Passivity_Condition} then becomes essential, because it eliminates roots that would otherwise correspond to exponentially growing solutions in a passive material.

            As in the Drude case (Fig.~\ref{fig:Number_of_Waves_Drude}), the conventional nondispersive regimes are recovered in the high-frequency limit, where the material response approaches a constant refractive index. In the Lorentz case, however, these conventional regimes are embedded within a substantially richer set of dispersion-mediated space-time states generated by the additional resonant branches. The Lorentz model therefore shows that resonance and loss do not merely perturb the nondispersive picture; they reorganize the scattering landscape itself by modifying the number, position and physical admissibility of the available channels.

\begin{figure*}
    \centering
    \includegraphics[width=\linewidth]{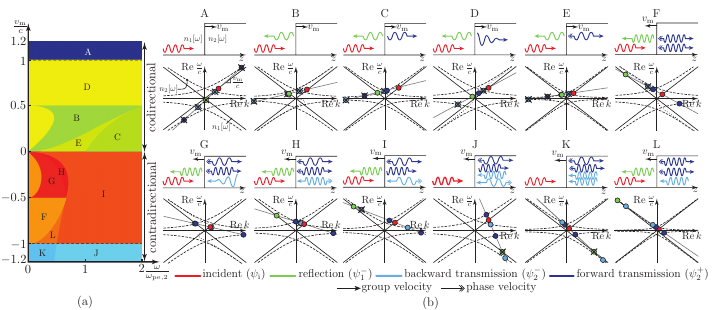}
    \caption{Dispersion-mediated space-time modes in a double-Drude medium [Eq.~\eqref{eq:Refractive_Index_Double_Drude_Media}], with $n_{\infty\text{e},1} = 1$, $\omega_{\text{pe},1} = 0$, $n_{\infty\text{m},1} = 1$, $\omega_{\text{pm},1}=0$ (vacuum) and $n_{\infty\text{e},2} = 1$, $\omega_{\text{pe},2} = 10$, $n_{\infty\text{m},2} = 1$ and $\omega_{\text{pm},2}=10$. (a)~Scattering regimes as a function of the modulation velocity, $v_{\text{m}}$, and the incident frequency, $\omega_{\text{i}}$. The dashed lines indicate the corresponding nondispersive velocity limits. (b)~Space-index representation of the different space-time states with associated spectral transition diagrams, where some of the solutions are crossed out because they do not satisfy the selection conditions in Eqs.~\eqref{eq:Selection_Conditions}. The dotted curves represent dispersion relations evaluated at fixed values of the imaginary part of the frequency solutions.}
    \label{fig:Number_of_Waves_Double_Drude}
\end{figure*}

        \subsubsection{Double-Drude Media}\label{subsubsec:Double-Drude_Media}
            \pati{Model Definition}{
            }

            As a third example, we consider a double-Drude model, in which both the permittivity and permeability are dispersive. This corresponds to a simultaneous modulation of the refractive index $n_{i}{\left[\omega\right]}$ and the wave impedance $\eta_{i}{\left[\omega\right]}$, given by~\cite{Veselago1968_Neg_Ref_PUB}
            \begin{subequations}\label{eq:Refractive_Index_Double_Drude_Media}
                \begin{align}
                n_{i}{\left[\omega\right]} &= \sqrt{\left(n_{\infty\text{e},i}^{2} - \frac{\omega_{\text{pe},i}^{2}}{\omega^{2}}\right)\left(n_{\infty\text{m},i}^{2} - \frac{\omega_{\text{pm},i}^{2}}{\omega^{2}}\right)}\,, \\
                \eta_{i}{\left[\omega\right]} &= \sqrt{\frac{n_{\infty\text{m},i}^{2} - \omega_{\text{pm},i}^{2}/\omega^{2}}{n_{\infty\text{e},i}^{2} - \omega_{\text{pe},i}^{2}/\omega^{2}}}\,,
                \end{align}
            \end{subequations}
            where~$n_{\infty\text{e},i}$ and~$n_{\infty\text{m},i}$ are the electric and magnetic high-frequency refractive indices, while~$\omega_{\text{pe},i}$ and~$\omega_{\text{pm},i}$ are the corresponding plasma frequencies of medium~\mbox{$i=1,2$}. A key feature of this model is the existence of frequency regions where both the permittivity and permeability are negative, leading to a negative refractive index. In such regions, the phase velocity and group velocity have opposite signs. Consequently, phase fronts and energy flow are antiparallel. In the spectral representation, this behavior is captured by a local slope (group velocity) that is opposite to the slope of the line connecting the point to the origin (phase velocity). Substituting Eq.~\eqref{eq:Refractive_Index_Double_Drude_Media} into Eqs.~\eqref{eq:General_Characteristic_Equations} again yields quartic equations (see Sec.~\ref{subsec:appendix:Double_Drude_Media}), producing up to four candidate solutions per medium. Figure~\ref{fig:Number_of_Waves_Double_Drude} shows the resulting scattering landscape for a configuration where the first medium is vacuum and the second medium is described by a double-Drude model [Eqs.~\eqref{eq:Refractive_Index_Double_Drude_Media}]. In this configuration, only waves in the second medium can enter the negative-index regime. The reflected waves in the first medium remain nondispersive and follow the standard Doppler relations (Sec.~\ref{subsec:appendix:Double_Drude_Media}).

            \pati{Single-Drude-Like Configurations}{
            }

            Several regimes of the single-Drude model (Fig.~\ref{fig:Number_of_Waves_Drude}) reappear in the double-Drude case. In particular, regions~A--D are recovered without modification, indicating that these regimes are robust with respect to the inclusion of magnetic dispersion. Region~E exhibits a topology similar to that of the subluminal regime (region~C), but with a crucial difference: the transmitted wave in the second medium lies within a negative-index region. In this case, the transition line couples to a dispersion branch with positive group velocity but negative phase velocity. As a result, the transmitted wave propagates forward in terms of energy transport, while its phase fronts propagate in the opposite direction. This decoupling of phase and energy flow has no counterpart in single-Drude media and leads to qualitatively different space-time field patterns.

            \pati{Pure Double-Drude Configurations}{
            }

            Beyond these single-Drude analogs, the double-Drude model supports a set of genuinely new scattering regimes (regions~F--L in Fig.~\ref{fig:Number_of_Waves_Double_Drude}) that have no counterpart in the single-Drude case (Fig.~\ref{fig:Number_of_Waves_Drude}). These regimes arise from the combined effect of electric and magnetic resonances, which introduce additional dispersion branches and allow for intersections with both positive- and negative-index modes. As a result, the number and type of admissible solutions vary strongly with both incident frequency and modulation velocity. A defining feature of these regimes is the presence of negative-index modes in the scattered field. Depending on the specific configuration, one or more waves may exhibit antiparallel phase and group velocities, leading to propagation directions that differ from those inferred from phase evolution alone. Most of these regimes occur for contradirectional interfaces, where the relative motion enhances coupling to the additional dispersion branches. However, region~F constitutes an exception, demonstrating that such modes are not exclusively tied to a specific interface direction. Overall, the double-Drude model illustrates how simultaneous electric and magnetic dispersion further enriches the space-time scattering landscape.

\begin{figure}
    \centering
    \includegraphics[width=0.95\linewidth]{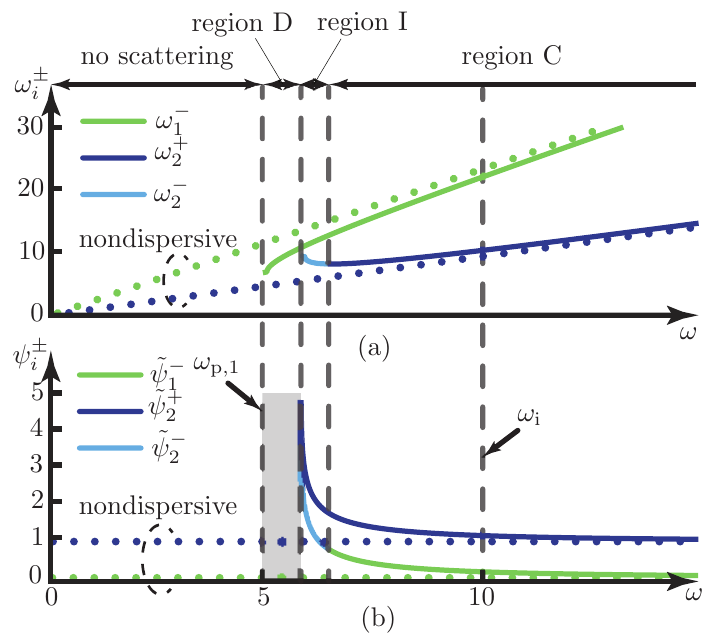}
    \caption{Application to the Drude dispersion model, with the same parameters as in Fig.~\ref{fig:Number_of_Waves_Drude}, and fixed modulation velocity, $v_{\text{m}}/c = -0.4$. (a)~Frequency transitions, determined by Eqs.~\eqref{eq:General_Characteristic_Equations}. (b)~Scattering amplitudes [Eqs.~\eqref{eq:Fourier_Scattered_Waves}]. The dashed line at $\omega = 10$ is the incident frequency used in the simulation of Fig.~\ref{fig:FDTD_Validation_Drude}.}
    \label{fig:Application_Drude_Scattering_Coefficient}
\end{figure}

\section{Scattering Coefficients}\label{sec:Scattering_Coefficients}
    \subsection{Theory}\label{subsec:Theory_Scat_Coef}
        \pati{Failure of Nondispersive Methods}{
        }
        
        Section~\ref{sec:Frequency_Transitions} focused exclusively on the spectral transitions of the scattered waves, obtained from the \emph{phase} continuity conditions [Eqs.~\eqref{eq:Phase_Matching_Conditions}]. However, a complete scattering description also requires the corresponding wave amplitudes, which follow from the \emph{boundary conditions} at the moving interface. For a constant-velocity interface, these impose the continuity of $E{\left[z,t\right]} - v_{\text{m}}B{\left[z,t\right]}$ and $H{\left[z,t\right]} - v_{\text{m}}D{\left[z,t\right]}$ at the interface $z=v_{\text{m}}t$~\cite{Caloz2019b_ST_Metamaterials_USTEM_PUB}. To apply these conditions, the induced fields $B{\left[z,t\right]}$ and $D{\left[z,t\right]}$ must be expressed in terms of the driving fields $H{\left[z,t\right]}$ and $E{\left[z,t\right]}$ through the constitutive relations of the medium. In \emph{nondispersive} media, the constitutive relations are instantaneous: $B{\left[z,t\right]} = n\eta/c\, H{\left[z,t\right]}$ and $D{\left[z,t\right]} = n/{\left(c\eta\right)}\,E{\left[z,t\right]}$, allowing direct substitution into the time-domain boundary conditions. The scattering amplitudes can then be obtained through straightforward algebra. In \emph{dispersive} media, by contrast, causality enforces a nonlocal temporal response: the induced fields depend on the entire past history of the excitation, so the constitutive relations take the form of time convolutions. As a result, the boundary conditions couple the field values at the interface to their full temporal evolution. A purely time-domain formulation thus becomes impractical, because evaluating the boundary conditions at a given time would require explicit knowledge of the fields at all prior times.
    
        \pati{Proposed Strategy}{
        }    

        This difficulty can be addressed by combining a time-domain description of the interface motion with a frequency-domain treatment of the material response (Sec.~\ref{sec:appendix:Scattering_Coefficients}). Starting from the boundary conditions in the spacetime domain, we evaluate them at the moving interface, $z = v_{\text{m}}t$. This eliminates the spatial coordinate from the problem, leaving boundary conditions that depend only on time. Next, we Fourier-transform these time-dependent boundary conditions into the frequency domain, where the dispersive constitutive relations of the medium become simple algebraic relations: $\tilde{B}{\left[z,\omega\right]} = n{\left[\omega\right]}\eta{\left[\omega\right]}/c \,\tilde{H}{\left[z,\omega\right]}$ and $\tilde{D}{\left[z,\omega\right]} = n{\left[\omega\right]}/{\left(c\eta{\left[\omega\right]}\right)}\,\tilde{E}{\left[z,\omega\right]}$. Each frequency component can then be treated independently, fully capturing the causal features of the medium. This mixed-domain approach separates the kinematics of the moving interface, naturally handled in the time domain, from the dispersive response of the medium, most conveniently treated in the frequency domain. As a result, the boundary conditions reduce to a closed set of algebraic equations for the spectral amplitudes of the scattered waves, eliminating the need to track the full temporal history of the fields.
                
        \pati{Results}{
        }

        As an example, we consider the configuration in which a single backward-propagating reflected wave in the first medium and a single forward-propagating transmitted wave in the second medium are generated, corresponding to the subluminal regime (region~C in Fig.~\ref{fig:Number_of_Waves_Drude}). For a monochromatic incident field with frequency $\omega_{\text{i}}$, the Fourier scattered waves are given by (Sec.~\ref{sec:appendix:Scattering_Coefficients})
        \begin{subequations}\label{eq:Fourier_Scattered_Waves}
            \begin{align}
                \tilde{\psi}_{1}^{-}\left[\omega\right] &= \frac{\eta_{1}{\left[\omega_{1}^{-}\right]}}{\eta_{1}{\left[\omega_{\text{i}}\right]}}\frac{\eta_{2}{\left[\omega_{2}^{+}\right]}-\eta_{1}{\left[\omega_{\text{i}}\right]}}{\eta_{2}{\left[\omega_{2}^{+}\right]}+\eta_{1}{\left[\omega_{1}^{-}\right]}} \nonumber\\
                &\hspace{2cm} \times\tilde{\psi}_{\text{i}}\left[\frac{1+n_{1}{\left[\omega_{1}^{-}\right]}v_{\text{m}}/c}{1- n_{1}{\left[\omega_{\text{i}}\right]}v_{\text{m}}/c}\omega\right]\,, \\
                \tilde{\psi}_{2}^{+}\left[\omega\right] &= \frac{\eta_{2}{\left[\omega_{2}^{+}\right]}}{\eta_{1}{\left[\omega_{\text{i}}\right]}}\frac{\eta_{1}{\left[\omega_{1}^{-}\right]}+\eta_{1}{\left[\omega_{\text{i}}\right]}}{\eta_{2}{\left[\omega_{2}^{+}\right]}+\eta_{1}{\left[\omega_{1}^{-}\right]}} \nonumber \\
                &\hspace{2cm} \times\tilde{\psi}_{\text{i}}\left[\frac{1-n_{2}{\left[\omega_{2}^{+}\right]}v_{\text{m}}/c}{1-n_{1}{\left[\omega_{\text{i}}\right]}v_{\text{m}}/c}\omega\right] \,.
            \end{align}
        \end{subequations}
        where $\omega_{1}^{-}$ and $\omega_{2}^{+}$ are the reflected and transmitted frequencies, respectively, determined by the phase-matching equations [Eqs.~\eqref{eq:General_Characteristic_Equations}]. The prefactors in Eqs.~\eqref{eq:Fourier_Scattered_Waves} describe the impedance mismatch between the two media, while the arguments of $\tilde{\psi}_{\text{i}}$ encode the inverse Doppler frequency shift induced by the interface motion. Although these expressions formally resemble the reflection and transmission coefficients of a nondispersive interface~\cite{Deck-Leger2019_Uni_Vel_USTEM_PUB}, the key distinction is that all quantities are evaluated at frequency-shifted arguments, making the coefficients inherently frequency dependent.
        
        \pati{Other Regions}{
        }

        Equations~\eqref{eq:Fourier_Scattered_Waves} only apply to the subluminal regime, where exactly one reflected and one transmitted wave are generated. Other regimes that also yield two scattered waves, such as region~I in Fig.~\ref{fig:Number_of_Waves_Drude}, can be treated using the same methodology, although the resulting expressions differ slightly because the participating scattered waves are of a different type. In regimes with more than two scattered waves, such as region~H in Fig.~\ref{fig:Number_of_Waves_Drude}, the system of equations becomes underdetermined and additional boundary conditions are required to fully determine the scattered amplitudes. This situation is analogous to nondispersive interluminal scattering, where an extra condition is also necessary in the contradirectional case~\cite{Ostrovskii1967_Inter_PUB,Li2026_Interluminal_USTEM}. We leave these cases for future work.

    \subsection{Application}\label{subsec:Application_Scat_Coef}
        \pati{Application to Drude Model}{
        }

        To gain physical insight into Eqs.~\eqref{eq:Fourier_Scattered_Waves}, we apply the formulation to a Drude dispersive medium [Eq.~\eqref{eq:Refractive_Index_Drude_Media}]. Figure~\ref{fig:Application_Drude_Scattering_Coefficient} plots the resulting frequency transitions (Fig.~\ref{fig:Application_Drude_Scattering_Coefficient}a) and Fourier scattering amplitudes (Fig.~\ref{fig:Application_Drude_Scattering_Coefficient}b) for a fixed modulation velocity~$v_{\text{m}}/c = -0.4$, corresponding to a horizontal slice in Fig.~\ref{fig:Number_of_Waves_Drude}a at~$v_{\text{m}}/c = -0.4$. For frequencies~$\omega < \omega_{\text{p},1}$, the incident wave lies below the plasma frequency and becomes evanescent in the first medium. As a result, no scattering with the interface occurs. For higher frequencies, region~D corresponds to interluminal scattering, where only a reflected wave is generated. The associated scattering amplitude is not captured by the present formulation and is therefore indicated in gray. This limitation reflects the broader difficulty of determining amplitudes in interluminal regimes, even in nondispersive systems~\cite{Li2026_Interluminal_USTEM}. Region~I represents a dispersion-mediated space-time mode, where multiple scattering channels coexist. In this regime, the scattering amplitudes exhibit strong enhancement and may appear to diverge, indicating that energy transfer is being transferred from the modulation to the wave~\cite{Li2025_Wave_Medium_Int_USTEM_PUB}. However, these amplitudes remain finite and bounded. Finally, in region~C (subluminal regime), the behavior simplifies. At sufficiently high frequencies, the dispersive effects weaken and both the frequency transitions and the scattering amplitudes asymptotically approach their nondispersive counterparts. This convergence provides a consistency check on the formulation and confirms that the proposed method correctly recovers the nondispersive limit.
        
\begin{figure}
    \centering
    \includegraphics[width=\linewidth]{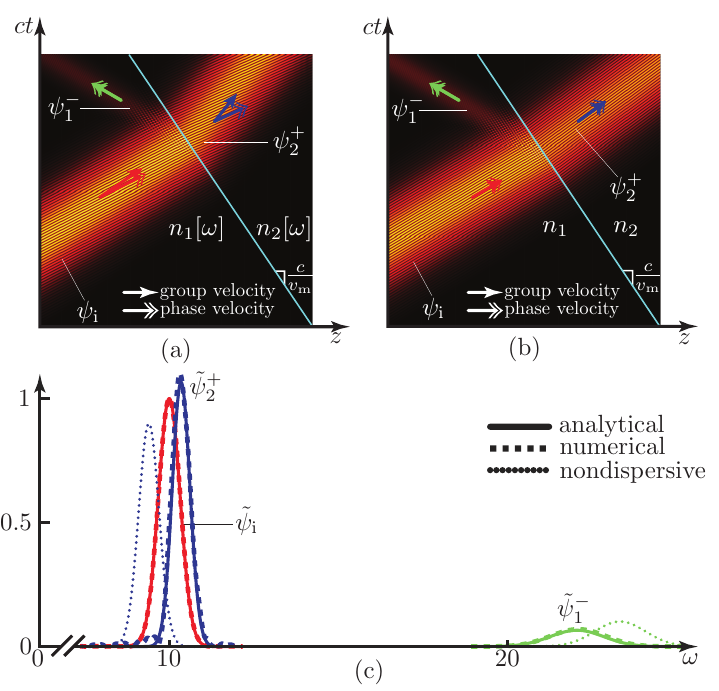}
    \caption{Comparison of scattering at a moving interface in dispersive and nondispersive media. (a)~Space-time diagram for the dispersive case at $\omega_{\text{i}} = 10$ and $v_{\text{m}}/c=-0.4$, with the same parameters as in Fig.~\ref{fig:Number_of_Waves_Drude}. (b)~Corresponding space-time diagram for the nondispersive case, with refractive indices equal to the high-frequency limits of~(a). (c)~Fourier spectra of the scattered fields: analytical predictions for both dispersive and nondispersive cases, along with numerical results obtained from a dispersive FDTD simulation.}
    \label{fig:FDTD_Validation_Drude}
\end{figure}

\section{Comparison with Nondispersive Media}\label{sec:Comparison_with_Nondispersive_Media}
    \pati{Comparison Dispersive and Nondispersive Scattering}{
    }
    
    We now compare the dispersive results with their nondispersive counterpart in the subluminal regime (region~C in Fig.~\ref{fig:Number_of_Waves_Drude}). Figure~\ref{fig:FDTD_Validation_Drude} shows electromagnetic scattering at a moving interface with velocity $v_{\text{m}}/c=-0.4$ and incident frequency $\omega_{\text{i}} = 10$, for both the Drude model [Eq.~\eqref{eq:Refractive_Index_Drude_Media}] and its nondispersive limit. The nondispersive reference is constructed by fixing the refractive indices to their high-frequency limits. This choice removes frequency dependence while preserving a consistent asymptotic behavior, ensuring that any observed differences arise solely from dispersion. Figures~\ref{fig:FDTD_Validation_Drude}a and \ref{fig:FDTD_Validation_Drude}b show the corresponding space-time diagrams for the dispersive and nondispersive cases, respectively, while Fig.~\ref{fig:FDTD_Validation_Drude}c presents the Fourier spectra of the incident, reflected and transmitted fields.

    Several fundamental differences emerge. First, the frequency transitions differ. In dispersive media, the phase-matching conditions depend on the frequency-dependent refractive index, leading to shifted reflected and transmitted frequencies relative to the nondispersive case, consistent with the trends observed in Fig.~\ref{fig:Application_Drude_Scattering_Coefficient}a. Second, the scattering amplitudes are altered. In nondispersive media, the reflection and transmission coefficients depend only on the constant impedances of the two media. In dispersive media, by contrast, the impedances are evaluated at the scattered frequencies, which themselves depend on the interface motion. The amplitudes are therefore coupled to the frequency transitions through the material response, consistent with Fig.~\ref{fig:Application_Drude_Scattering_Coefficient}b. Third, dispersion separates phase and group velocities. In the nondispersive case, $v_{\text{g},i}=v_{\text{p},i}$, whereas in the dispersive case $v_{\text{g},i}\neq v_{\text{p},i}$. This difference is visible in Fig.~\ref{fig:FDTD_Validation_Drude}a, where the slopes of the phase fronts and of the pulse envelopes no longer coincide. 
    
    \pati{FDTD Validation}{
    }

    Finally, the analytical predictions are validated using a finite-difference time-domain simulation based on a generalized Yee grid~\cite{Deck-Leger2022_FDTD_USTEM_PUB,Bahrami2023_FDTD_ASTEM_PUB}, which explicitly enforces the moving-interface boundary conditions. Excellent agreement is observed between the analytical results [Eqs.~\eqref{eq:Fourier_Scattered_Waves}] and the numerical simulations, as shown in Fig.~\ref{fig:FDTD_Validation_Drude}c. This confirms that the proposed mixed-domain framework accurately captures scattering at moving interfaces in dispersive media and correctly accounts for both the frequency transitions and the associated amplitude variations.

\section{Experimental Outlook}\label{sec:Experimental_Outlook}
    \pati{General Comment}{
    }

    The dispersion-mediated space-time modes introduced in this work rely on the realization of sharp, at least subwavelength and subperiod-wide~\footnote{If the interface transition length and transition time, which are necessarily finite in practice because of causality, are respectively much smaller than the wavelength and shorter than the period of the incident wave, then the scattering response will be essentially identical to that computed for an ideal, step-like interface~\cite{Caloz2022_GSTEMs_ASTEM_PUB,Li2025_Graded_Index_USTEM_PUB}. Moreover, if only the transmitted waves, corresponding to the right-directed arrows in the second medium in Fig.~\ref{fig:Number_of_Waves_Drude}b, are required, an adiabatic (larger and slower) interface transition may still produce the predicted waves.}, moving interfaces. Realizing such interfaces may pose experimental challenges, since methods for generating space-time boundaries with arbitrary control are not yet fully established. Nevertheless, recent advances in the implementation of time-varying media~\cite{Alu2023_Temp_Refl_TEM_PUB,Alu2023_Coherent_Wave_Control_TEM_PUB,Peroulis2024_Time_Ref_TEM_PUB,Boyd2020_Time_Refr_ENZ_TEM_PUB,Segev2023_Single_Cycle_TEM_PUB,Kinsey2025_ST_Knife_TEM_PUB} provide a promising foundation for the development of space-time-varying systems. In the following, we outline several existing experimental platforms that could be adapted to investigate the phenomena predicted in this paper.

    \pati{Microwave Regime with Switches}{
    }

    In the microwave regime, temporal interfaces have been experimentally realized through rapid modulation of the effective impedance of artificial transmission lines~\cite{Alu2023_Temp_Refl_TEM_PUB,Alu2023_Coherent_Wave_Control_TEM_PUB,Peroulis2024_Time_Ref_TEM_PUB}. Such modulations are typically achieved using high-speed electronic switches or photodiodes that abruptly alter the electrical properties of the transmission line. Extending these concepts to space-time varying systems requires the switching process to be coordinated both temporally and spatially. One possible approach is to employ a field-programmable gate array (FPGA) to drive a sequence of switches or photodiodes with precisely controlled delays. By synchronizing the activation of individual switching elements along the transmission line, an effective moving interface can be synthesized. Alternative microwave implementations may exploit corporate feed networks combined with tailored optical delay lines to generate the required moving switching pattern. The dispersive character required for the observation of the modes presented here can be introduced by employing transmission lines fabricated on strongly dispersive substrates, such as lithium niobate, or artificial transmission lines~\cite{Caloz2006_TL_BOOK}.

    \pati{Optical Regime with ENZ}{
    }

    In the optical regime, significant progress has recently been achieved using epsilon-near-zero (ENZ) materials in pump-probe experiments, where ultrafast optical excitation produces temporal changes in the refractive index~\cite{Boyd2020_Time_Refr_ENZ_TEM_PUB,Segev2023_Single_Cycle_TEM_PUB,Kinsey2025_ST_Knife_TEM_PUB}. These demonstrations provide a natural starting point for realizing space-time varying refractive-index fronts at optical frequencies. One possible extension consists of illuminating the material with an \emph{obliquely} incident pump beam. In this configuration, the pump wavefront reaches different spatial locations at different times, thereby creating an effective refractive-index front that propagates across the medium with a controllable velocity in both the subluminal and superluminal regimes. Since the optical response of ENZ materials is commonly described by a Drude-type dispersion model, the theoretical framework developed in Sec.~\ref{subsubsec:Drude_Media} is directly applicable to such systems, making them particularly attractive candidates for experimental validation.

 \pati{Optical Regime with Nonlinear Fibers}{
    }
    
    A second optics-compatible platform is based on nonlinear dispersive waveguides, such as optical fibers~\cite{Agrawal2024_Soliton_Raman_USTEM_PUB,Agrawal2015_Temporal_Analog_USTEM_PUB,Agrawal2023_Exp_Temp_Wavegui_USTEM_PUB}. In these systems, moving refractive-index boundaries can be generated through the Kerr effect by launching an intense pump pulse (soliton) into the waveguide. The pump pulse induces a localized refractive-index perturbation that propagates with the pulse, thereby forming an effective moving interface. A weaker probe pulse then interacts with this moving boundary. Taken together, these microwave and optical platforms suggest that the dispersion-mediated space-time modes identified in this work are experimentally accessible with foreseeable extensions of existing technologies.

\section{Conclusions}\label{sec:Conclusions}
    \pati{Conclusions}{
    }

    We developed a general theory for electromagnetic scattering at moving interfaces between dispersive media. The theory determines the allowed frequency transitions from phase continuity and selects the physically admissible scattered waves by combining a group-velocity separation condition with a media passivity constraint. Application to Drude, Lorentz and double-Drude media showed that dispersion qualitatively reshapes the space-time scattering landscape. Beyond the conventional nondispersive regimes, dispersive media support additional scattering configurations that have no nondispersive counterpart. We further derived closed-form Fourier-domain scattering amplitudes for the two-wave class using a mixed-domain formulation that combines time-domain interface kinematics with frequency-domain constitutive relations. These amplitudes recover the nondispersive limit at high frequencies and are in agreement with dispersive FDTD simulations. The present work therefore establishes a unified kinematic framework for dispersive space-time scattering.

    \pati{Future Work}{
    }
    
     A first direction for future research is the extension of the amplitude theory to multi-wave and interluminal regimes, where additional boundary conditions are required. A second direction is to apply the methodology to accelerated and multi-velocity systems. Because the approach developed here does not rely on frame hopping, it can, in principle, be adapted to such systems in dispersive media, including phenomena such as space-time focusing, originally introduced by Ostrovski\u{\i}~\cite{Ostrovskii1975_Lens_ASTEM_PUB,Ostrovskii1971_Geometrics_Disp_ASTEM_PUB}, and the recently proposed space-time wedges~\cite{Bahrami2025_Wedges_USTEM_PUB}.

    \textbf{Acknowledgments} \\
    K.D.K. is supported by the Research Foundation -- Flanders (FWO) doctoral fellowship 1174526N.

\appendix
\section{Characteristic Equations}

\subsection{Drude Media}\label{subsec:appendix:Drude_Media}
    \pati{Drude Media}{
    }

    In this appendix, we derive the characteristic equations for several dispersive models and their corresponding frequency transitions, starting with the Drude model. The refractive index profile of the Drude medium is given in Eq.~\eqref{eq:Refractive_Index_Drude_Media}. To obtain the frequency relations at a moving interface, we first rewrite Eqs.~\eqref{eq:General_Characteristic_Equations} as
    \begin{subequations}\label{eq:appendix:Frequency_Transitions_Rewritten}
        \begin{align}
            \omega_{1}n_{1}{\left[\omega_{1}\right]} &= - \frac{\omega_{1}-\left(1-n_{1}{\left[\omega_{\text{i}}\right]}v_{\text{m}}/c\right)\omega_{\text{i}}}{v_{\text{m}}/c}\,, \\
            \omega_{2}n_{2}{\left[\omega_{2}\right]} &= \phantom{+}\frac{\omega_{2}-\left(1-n_{1}{\left[\omega_{\text{i}}\right]}v_{\text{m}}/c\right)\omega_{\text{i}}}{v_{\text{m}}/c}\,.
        \end{align}
    \end{subequations}
    Squaring both sides of Eqs.~\eqref{eq:appendix:Frequency_Transitions_Rewritten} and substituting Eq.~\eqref{eq:Refractive_Index_Drude_Media} yields the characteristic equations for medium $i=1,2$ as
    \begin{equation}\label{eq:appendix:Characteristic_Equations_Drude_Media}
            \Omega_{i}\omega_{i}^{2} -2\Gamma\omega_{i} + \frac{v_{\text{m}}^{2}}{c^{2}}\omega_{\text{p},i}^{2} + \Gamma^{2} = 0 \,.
    \end{equation}
    where the parameters $\Omega_{i}$ and $\Gamma$ are defined as
    \begin{subequations}\label{eq:appendix:Definition_Omega_and_Gamma}
        \begin{align}
            \Omega_{i} &= 1-n_{\infty,i}^{2}\frac{v_{\text{m}}^{2}}{c^{2}}\,, \label{eq:appendix:Definition_Omega} \\
            \Gamma &= \left(1-n_{1}{\left[\omega_{\text{i}}\right]}\frac{v_{\text{m}}}{c}\right)\omega_{\text{i}}\,. \label{eq:appendix:Definition_Gamma}
        \end{align}
    \end{subequations}
    Equation~\eqref{eq:appendix:Characteristic_Equations_Drude_Media} is a quadratic polynomial in $\omega_{i}$, which can be easily solved as
    \begin{equation}\label{eq:Frequency_Transitions_Example}
        \omega_{i} = \frac{1}{\Omega_{i}}\left(\Gamma \pm \frac{v_{\text{m}}}{c}\sqrt{n_{\infty,i}^{2}\Gamma^{2}-\Omega_{i}\omega_{\text{p},i}^{2}}\right) \,. 
    \end{equation}

\subsection{Lorentz Media}\label{subsec:appendix:Lorentz_Media}
    \pati{Lorentz Media}{
    }

    For Lorentz media, the refractive index is given by Eq.~\eqref{eq:Refractive_Index_Lorentz_Media}. Inserting Eq.~\eqref{eq:Refractive_Index_Lorentz_Media} into Eqs.~\eqref{eq:appendix:Frequency_Transitions_Rewritten} and squaring both sides results in a quartic polynomial in $\omega_{i}$:
    \begin{equation}\label{eq:appendix:Characteristic_Equations_Lorentz_Media}
        \begin{split}
            &\Omega_{i}\omega_{i}^{4} - \left(2\Gamma - i\gamma_{i}\Omega_{i}\right)\omega_{i}^{3} \\
            &\hspace{1cm}- \left(\Omega_{i}\omega_{0,i}^{2} + 2i\gamma_{i}\Gamma -\frac{v_{\text{m}}^{2}}{c^{2}}\omega_{\text{p},i}^{2} - \Gamma^{2}\right)\omega_{i}^{2}  \\
            &\hspace{2cm}+ \Gamma\left(2\omega_{0,i}^{2} + i\gamma_{i}\Gamma\right)\omega_{i} - \Gamma^{2}\omega_{0,i}^{2} = 0 \,,
        \end{split}
    \end{equation}
    where $\Omega_{i}$ and $\Gamma$ are defined in Eqs.~\eqref{eq:appendix:Definition_Omega_and_Gamma}. Note that the Drude case [Eq.~\eqref{eq:appendix:Characteristic_Equations_Drude_Media}] is recovered by setting $\omega_{0,i} = 0 = \gamma_{i}$ in Eq.~\eqref{eq:appendix:Characteristic_Equations_Lorentz_Media}. 

\subsection{Double-Drude Media}\label{subsec:appendix:Double_Drude_Media}
    \pati{Double-Drude Media}{
    }

    For a double-Drude medium with refractive-index profile given in Eq.~\eqref{eq:Refractive_Index_Double_Drude_Media}, substitution into Eqs.~\eqref{eq:appendix:Frequency_Transitions_Rewritten} and squaring both sides yields
    \begin{equation}\label{eq:appendix:Characteristic_Equations_Double_Drude_Media}
        \begin{split}
            &\Omega_{\text{em},i}\omega_{i}^{4} - 2\Gamma\omega_{i}^{3}  \\
            &\hspace{0.5cm}+\left(\Gamma^{2}+\left(n_{\infty\text{m},i}^{2}\omega_{\text{pe},i}^{2}+n_{\infty\text{e},i}^{2}\omega_{\text{pm},i}^{2}\right)\frac{v_{\text{m}}^{2}}{c^{2}}\right)\omega_{i}^{2} \\
            &\hspace{1.5cm}- \omega_{\text{pe},i}^{2}\omega_{\text{pm},i}^{2}\frac{v_{\text{m}}^{2}}{c^{2}} = 0\,,
        \end{split}
    \end{equation}
    where $\Gamma$ is defined in Eq.~\eqref{eq:appendix:Definition_Gamma} and
    \begin{equation}
        \Omega_{\text{em},i} = 1-n_{\infty\text{e},i}^{2}n_{\infty\text{m},i}^{2}\frac{v_{\text{m}}^{2}}{c^{2}}\,.
    \end{equation}
    Equation~\eqref{eq:appendix:Characteristic_Equations_Double_Drude_Media} is a quartic polynomial in $\omega_{i}$. For the particular case $n_{\infty\text{e},1} = 1 = n_{\infty\text{m},1}$, $\omega_{\text{pe},1} = 0 = \omega_{\text{pm},1}$ (vacuum) and $n_{\infty\text{e},2} = 1 = n_{\infty\text{m},2}$, $\omega_{\text{pe},2} = \omega_{\text{pm},2}$ (Fig.~\ref{fig:Number_of_Waves_Double_Drude}), Eq.~\eqref{eq:appendix:Characteristic_Equations_Double_Drude_Media} can be solved analytically as
    \begin{subequations}
        \begin{align}
            \omega_{1} &= \frac{1-v_{\text{m}}/c}{1+v_{\text{m}}/c}\omega_{\text{i}}\,, \label{eq:appendix:omega_1_Double_Drude} \\
            \omega_{2} &= \frac{1}{2\left(1+\sigma v_{\text{m}}/c\right)} \nonumber\\
            &\hspace{0.5cm}\times\Bigg(\Gamma \pm \sqrt{\Gamma^{2}+4\sigma\left(1+\sigma\frac{v_{\text{m}}}{c}\right)\frac{v_{\text{m}}}{c}\omega_{\text{pe},2}^{2}}\Bigg)\,,
        \end{align}
    \end{subequations}
    where $n_{1}{\left[\omega\right]}=1$ for the chosen parameters and $\sigma=\pm 1$, yielding four possible solutions for $\omega_{2}$. Equation~\eqref{eq:appendix:omega_1_Double_Drude} recovers the standard reflection Doppler shift for a moving interface~\cite{Caloz2019b_ST_Metamaterials_USTEM_PUB}.

\section{Scattering Coefficients}\label{sec:appendix:Scattering_Coefficients}
    \pati{Assumptions}{
    }

    This section derives the electromagnetic scattering solutions at a moving interface with constant velocity $v_{\text{m}}$. We consider the case where an incident wave, $\psi_{\text{i}}$, generates one backward-propagating reflection in the first medium, $\psi_{1}^{-}$, and one forward-propagating transmission in the second medium, $\psi_{2}^{+}$ (subluminal regime, e.g., region~C in Fig.~\ref{fig:Number_of_Waves_Drude}). The interface separates two isotropic, linear, dispersive media, characterized by frequency-dependent refractive indices $n_{i}{\left[\omega\right]}$ and impedances $\eta_{i}{\left[\omega\right]}$, with $i=1,2$ denoting the medium. We consider a one-dimensional geometry, in which the interface moves along the $z$-direction, the electric field is polarized along the $x$-direction and the magnetic field along the $y$-direction. The interface follows the trajectory $z{\left[t\right]} = v_{\text{m}}t$.
    
    \pati{Traveling Wave Form}{
    }

    A forward-propagating monochromatic incident wave with frequency $\omega_{\text{i}}$ in the first medium can be written as
    \begin{equation}\label{eq:appendix:General_Forward_Incident_Wave}
        \exp\left(i\left(k_{1}{\left[\omega_{\text{i}}\right]}z-\omega_{\text{i}}t\right)\right) = \exp\left(i\omega_{\text{i}}\left(n_{1}{\left[\omega_{\text{i}}\right]}\frac{z}{c}-t\right)\right)\,,
    \end{equation}
    where $k_{1}{\left[\omega_{\text{i}}\right]} = n_{1}{\left[\omega_{\text{i}}\right]}\omega_{\text{i}}/c$ [Eq.~\eqref{eq:Dispersion_Relation}]. Equation~\eqref{eq:appendix:General_Forward_Incident_Wave} defines a natural traveling-wave variable $\phi = n_{1}{\left[\omega_{\text{i}}\right]}z/c-t$, with phase velocity $v_{\text{p}} = c/n_{1}{\left[\omega_{\text{i}}\right]}$. Therefore, we propose the ansatz that all scattered waves can be expressed in the general form $\psi_{i}^{\pm}{\left[\phi_{i}^{\pm}\right]} = \exp\left(i\omega_{i}^{\pm}\phi_{i}^{\pm}\right)$, where $\phi_{i}^{\pm} = \pm\left(n_{i}{\left[\omega_{i}^{\pm}\right]}z/c\mp t\right)$, $i$ denotes the medium and the superscript denotes forward ($+$) propagating waves and backward ($-$) propagating waves. 
    
    \pati{General Traveling Waves Subluminal}{
    }

    The electric fields may be written within the ansatz as
    \begin{subequations}\label{eq:appendix:Electric_Waveforms}
        \begin{align}
            E_{\text{i}} &= \psi_{\text{i}}{\left[n_{1}{\left[\omega_{\text{i}}\right]}\frac{z}{c}-t\right]} \, , \\
            E_{1}^{-}    &= \psi_{1}^{-}{\left[-\left(n_{1}{\left[\omega_{1}^{-}\right]}\frac{z}{c}+t\right)\right]}\,, \\
            E_{2}^{+}    &= \psi_{2}^{+}{\left[n_{2}{\left[\omega_{2}^{+}\right]}\frac{z}{c}-t\right]}\,, 
        \end{align}
    \end{subequations}
    and the magnetic fields can be written as
    \begin{subequations}\label{eq:appendix:Magnetic_Waveforms}
        \begin{align}
            H_{\text{i}} &=  \frac{1}{\eta_{1}{\left[\omega_{\text{i}}\right]}}\psi_{\text{i}}{\left[n_{1}{\left[\omega_{\text{i}}\right]}\frac{z}{c}-t\right]} \, , \\
            H_{1}^{-}    &= -\frac{1}{\eta_{1}{\left[\omega_{1}^{-}\right]}}\psi_{1}^{-}{\left[-\left(n_{1}{\left[\omega_{1}^{-}\right]}\frac{z}{c}+t\right)\right]}\,, \\
            H_{2}^{+}    &=  \frac{1}{\eta_{2}{\left[\omega_{2}^{+}\right]}}\psi_{2}^{+}{\left[n_{2}{\left[\omega_{2}^{+}\right]}\frac{z}{c}-t\right]}\,, 
        \end{align}
    \end{subequations}
    where $\psi_{\text{i}}$, $\psi_{1}^{-}$ and $\psi_{2}^{+}$ denote the incident, reflected and transmitted waves, respectively, and $\omega_{\text{i}}$, $\omega_{1}^{-}$ and $\omega_{2}^{+}$ are the corresponding frequencies. The reflected and transmitted frequencies are determined by the phase-matching conditions [Eqs.~\eqref{eq:General_Characteristic_Equations}] or by Eq.~\eqref{eq:Frequency_Transitions_Example} for the Drude-model example [Eq.~\eqref{eq:Refractive_Index_Drude_Media}].

    \pati{Boundary Conditions}{
    }

    The time-domain boundary conditions at $z=v_{\text{m}}t$ are~\cite{Caloz2019b_ST_Metamaterials_USTEM_PUB}
    \begin{widetext}
        \begin{subequations}
            \begin{align}
                &\left.\left(E_{\text{i}}{\left[n_{1}{\left[\omega_{\text{i}}\right]}\frac{z}{c} - t\right]} - v_{\text{m}}B_{\text{i}}{\left[n_{1}{\left[\omega_{\text{i}}\right]}\frac{z}{c} - t\right]}\right) + \left(E_{1}^{-}{\left[-\left(n_{1}{\left[\omega_{1}^{-}\right]}\frac{z}{c} + t\right)\right]} - v_{\text{m}}B_{1}^{-}{\left[-\left(n_{1}{\left[\omega_{1}^{-}\right]}\frac{z}{c} + t\right)\right]}\right)\right|_{z=v_{\text{m}}t} \nonumber\\
                &\hspace{8cm}= \left.\left(E_{2}^{+}{\left[n_{2}{\left[\omega_{2}^{+}\right]}\frac{z}{c} - t\right]} - v_{\text{m}}B_{2}^{+}{\left[n_{2}{\left[\omega_{2}^{+}\right]}\frac{z}{c} - t\right]}\right)\right|_{z=v_{\text{m}}t} \, , \\
                &\left.\left(H_{\text{i}}{\left[n_{1}{\left[\omega_{\text{i}}\right]}\frac{z}{c} - t\right]} - v_{\text{m}}D_{\text{i}}{\left[n_{1}{\left[\omega_{\text{i}}\right]}\frac{z}{c} - t\right]}\right) + \left(H_{1}^{-}{\left[-\left(n_{1}{\left[\omega_{1}^{-}\right]}\frac{z}{c} + t\right)\right]} - v_{\text{m}}D_{1}^{-}{\left[-\left(n_{1}{\left[\omega_{1}^{-}\right]}\frac{z}{c} + t\right)\right]}\right)\right|_{z=v_{\text{m}}t} \nonumber \\
                &\hspace{8cm}= \left.\left(H_{2}^{+}{\left[n_{2}{\left[\omega_{2}^{+}\right]}\frac{z}{c} - t\right]} - v_{\text{m}}D_{2}^{+}{\left[n_{2}{\left[\omega_{2}^{+}\right]}\frac{z}{c} - t\right]}\right)\right|_{z=v_{\text{m}}t} \,,
            \end{align}
        \end{subequations}
    \end{widetext}
    where all fields are evaluated at the interface position. Incorporating the interface trajectory yields
    \begin{widetext}        
        \begin{subequations}\label{eq:appendix:System_of_Equations_Subluminal}
            \begin{align}
                &E_{\text{i}}{\left[-\left(1 - n_{1}{\left[\omega_{\text{i}}\right]}\frac{v_{\text{m}}}{c}\right)t\right]} - v_{\text{m}}B_{\text{i}}{\left[-\left(1 - n_{1}{\left[\omega_{\text{i}}\right]}\frac{v_{\text{m}}}{c}\right)t\right]} + E_{1}^{-}{\left[-\left(1+n_{1}{\left[\omega_{1}^{-}\right]}\frac{v_{\text{m}}}{c}\right)t\right]} - v_{\text{m}}B_{1}^{-}{\left[-\left(1 + n_{1}{\left[\omega_{1}^{-}\right]}\frac{v_{\text{m}}}{c}\right)t\right]} \nonumber \\
                &\hspace{7cm}= E_{2}^{+}{\left[-\left(1 - n_{2}{\left[\omega_{2}^{+}\right]}\frac{v_{\text{m}}}{c}\right)t\right]} - v_{\text{m}}B_{2}^{+}{\left[-\left(1 - n_{2}{\left[\omega_{2}^{+}\right]}\frac{v_{\text{m}}}{c}\right)t\right]}\,, \\
                &H_{\text{i}}{\left[-\left(1 - n_{1}{\left[\omega_{\text{i}}\right]}\frac{v_{\text{m}}}{c}\right)t\right]} - v_{\text{m}}D_{\text{i}}{\left[-\left(1 - n_{1}{\left[\omega_{\text{i}}\right]}\frac{v_{\text{m}}}{c}\right)t\right]} + H_{1}^{-}{\left[-\left(1 + n_{1}{\left[\omega_{1}^{-}\right]}\frac{v_{\text{m}}}{c}\right)t\right]} - v_{\text{m}}D_{1}^{-}{\left[-\left(1 + n_{1}{\left[\omega_{1}^{-}\right]}\frac{v_{\text{m}}}{c}\right)t\right]} \nonumber \\
                &\hspace{7cm}= H_{2}^{+}{\left[-\left(1 - n_{2}{\left[\omega_{2}^{+}\right]}\frac{v_{\text{m}}}{c}\right)t\right]} - v_{\text{m}}D_{2}^{+}{\left[-\left(1 - n_{2}{\left[\omega_{2}^{+}\right]}\frac{v_{\text{m}}}{c}\right)t\right]}\,.
            \end{align}
        \end{subequations}
    \end{widetext}
    Now comes the central idea of the proposed method. We Fourier-transform Eqs.~\eqref{eq:appendix:System_of_Equations_Subluminal}, while including the frequency-domain constitutive relations, inserting the general waveforms Eqs.~\eqref{eq:appendix:Electric_Waveforms} and \eqref{eq:appendix:Magnetic_Waveforms}, and using the Fourier scaling property
    \begin{equation}
        \mathcal{F}\left\{f{\left[at\right]}\right\}{\left[\omega\right]} = \frac{1}{\left|a\right|}\tilde{f}{\left[\frac{\omega}{a}\right]}\,.
    \end{equation}
    This leads to the following simple system of equations for the reflected and transmitted waves:
    \begin{widetext}
        \begin{subequations}\label{eq:appendix:Fourier_System_of_Equations_Subluminal}
            \begin{align}
                \tilde{\psi}_{\text{i}}\left[-\frac{\omega}{1- n_{1}{\left[\omega_{\text{i}}\right]}v_{\text{m}}/c}\right] + \tilde{\psi}_{1}^{-}\left[-\frac{\omega}{1+ n_{1}{\left[\omega_{1}^{-}\right]}v_{\text{m}}/c}\right] &= \tilde{\psi}_{2}^{+}\left[-\frac{\omega}{1 - n_{2}{\left[\omega_{2}^{+}\right]}v_{\text{m}}/c}\right] \,, \\
                \frac{1}{\eta_{1}{\left[\omega_{\text{i}}\right]}}\tilde{\psi}_{\text{i}}\left[-\frac{\omega}{1- n_{1}{\left[\omega_{\text{i}}\right]}v_{\text{m}}/c}\right] -\frac{1}{\eta_{1}{\left[\omega_{1}^{-}\right]}}\tilde{\psi}_{1}^{-}\left[-\frac{\omega}{1 + n_{1}{\left[\omega_{1}^{-}\right]}v_{\text{m}}/c}\right] &= \frac{1}{\eta_{2}{\left[\omega_{2}^{+}\right]}}\tilde{\psi}_{2}^{+}\left[-\frac{\omega}{1 - n_{2}{\left[\omega_{2}^{+}\right]}v_{\text{m}}/c}\right]\,.
            \end{align}
        \end{subequations}
    \end{widetext}
    Solving Eqs.~\eqref{eq:appendix:Fourier_System_of_Equations_Subluminal} for the reflected and transmitted waves yields
    \begin{subequations}\label{eq:appendix:Solution_Fourier_System_of_Equations_Subluminal}
        \begin{align}
            \tilde{\psi}_{1}^{-}\left[-\frac{\omega}{1 + n_{1}{\left[\omega_{1}^{-}\right]}v_{\text{m}}/c}\right] &= \frac{\eta_{1}{\left[\omega_{1}^{-}\right]}}{\eta_{1}{\left[\omega_{\text{i}}\right]}}\frac{\eta_{2}{\left[\omega_{2}^{+}\right]}-\eta_{1}{\left[\omega_{\text{i}}\right]}}{\eta_{2}{\left[\omega_{2}^{+}\right]}+\eta_{1}{\left[\omega_{1}^{-}\right]}} \nonumber \\
            &\times\tilde{\psi}_{\text{i}}\left[-\frac{\omega}{1 - n_{1}{\left[\omega_{\text{i}}\right]}v_{\text{m}}/c}\right]\,, \\
            \tilde{\psi}_{2}^{+}\left[-\frac{\omega}{1 - n_{2}{\left[\omega_{2}^{+}\right]}v_{\text{m}}/c}\right] &= \frac{\eta_{2}{\left[\omega_{2}^{+}\right]}}{\eta_{1}{\left[\omega_{\text{i}}\right]}}\frac{\eta_{1}{\left[\omega_{1}^{-}\right]}+\eta_{1}{\left[\omega_{\text{i}}\right]}}{\eta_{1}{\left[\omega_{1}^{-}\right]}+\eta_{2}{\left[\omega_{2}^{+}\right]}} \nonumber \\
            &\times\tilde{\psi}_{\text{i}}\left[-\frac{\omega}{1 - n_{1}{\left[\omega_{\text{i}}\right]}v_{\text{m}}/c}\right]\,.
        \end{align}
    \end{subequations}
    To express the scattered fields again as functions of the usual frequency variable $\omega$, we finally introduce spectral variables that absorb the Doppler prefactors appearing in Eqs.~\eqref{eq:appendix:Solution_Fourier_System_of_Equations_Subluminal},
    \begin{subequations}\label{eq:appendix:Change_of_Variables_Subluminal}
        \begin{align}
            \omega &\mapsto -\left(1 + n_{1}{\left[\omega_{1}^{-}\right]}\frac{v_{\text{m}}}{c}\right)\omega\,, \\
            \omega &\mapsto -\left(1 - n_{2}{\left[\omega_{2}^{+}\right]}\frac{v_{\text{m}}}{c}\right)\omega\,.
        \end{align}
    \end{subequations}
    Inserting Eqs.~\eqref{eq:appendix:Change_of_Variables_Subluminal} into Eqs.~\eqref{eq:appendix:Solution_Fourier_System_of_Equations_Subluminal} yields the final solutions
    \begin{subequations}
        \begin{align}
            \tilde{\psi}_{1}^{-}\left[\omega\right]& = \frac{\eta_{1}{\left[\omega_{1}^{-}\right]}}{\eta_{1}{\left[\omega_{\text{i}}\right]}}\frac{\eta_{2}{\left[\omega_{2}^{+}\right]}-\eta_{1}{\left[\omega_{\text{i}}\right]}}{\eta_{2}{\left[\omega_{2}^{+}\right]}+\eta_{1}{\left[\omega_{1}^{-}\right]}} \nonumber \\
            &\hspace{2cm}\times\tilde{\psi}_{\text{i}}\left[\frac{1+n_{1}{\left[\omega_{1}^{-}\right]}v_{\text{m}}/c}{1- n_{1}{\left[\omega_{\text{i}}\right]}v_{\text{m}}/c}\omega\right]\,, \\
            \tilde{\psi}_{2}^{+}\left[\omega\right] &= \frac{\eta_{2}{\left[\omega_{2}^{+}\right]}}{\eta_{1}{\left[\omega_{\text{i}}\right]}}\frac{\eta_{1}{\left[\omega_{1}^{-}\right]}+\eta_{1}{\left[\omega_{\text{i}}\right]}}{\eta_{2}{\left[\omega_{2}^{+}\right]}+\eta_{1}{\left[\omega_{1}^{-}\right]}} \nonumber \\
            &\hspace{2cm}\times\tilde{\psi}_{\text{i}}\left[\frac{1-n_{2}{\left[\omega_{2}^{+}\right]}v_{\text{m}}/c}{1-n_{1}{\left[\omega_{\text{i}}\right]}v_{\text{m}}/c}\omega\right]\,.
        \end{align}
    \end{subequations}
        
\bibliography{main}
\end{document}